\documentclass[aps,prd,preprint,superscriptaddress,longbibliography]{revtex4-1}
\usepackage{hyphenat}

\usepackage[T1]{fontenc}
\usepackage[utf8]{inputenc}
\usepackage{lmodern}
\usepackage{amsmath,amssymb,mathtools,bm}
\usepackage{booktabs}
\usepackage{graphicx}
\usepackage{microtype}
\usepackage{xcolor}
\usepackage[normalem]{ulem}

\newcommand{\ii}{\mathrm{i}}
\newcommand{\ee}{\mathrm{e}}

\newcommand{\Tr}{\operatorname{Tr}}

\newcommand{\Hil}{\mathcal H}

\newcommand{\Vhat}{\hat V}
\newcommand{\Uhat}{\hat U}

\usepackage[colorlinks=true,linkcolor=blue,citecolor=blue,urlcolor=blue]{hyperref}

\begin{document}

\title{Charged-Lepton Koide Geometry from a Green-Dressed Compact Family Cycle}

\author{Kirill Shulga}
\email{kirill\_shulga@protonmail.ch}
\affiliation{International Center for Elementary Particle Physics, The University of Tokyo, Tokyo 113-0033, Japan}

\date{\today}

\begin{abstract}

Koide's charged-lepton relation suggests that $(\sqrt{m_e},\sqrt{m_\mu},\sqrt{m_\tau})$ is the natural family vector. We construct an effective compact-cycle realization in which this vector is sampled from a real amplitude $Z(\varphi)$ on an internal circle, while the masses arise from quadratic chiral-profile overlaps. The coherent branch-real rank-one sector is an explicit assumption of the effective theory and realizes the equal singlet--doublet norm associated with Koide's $45^\circ$ cone. The empirical proximity of the $Z_3$ phase to $2/9$ predates the present work. The new ingredient proposed here is an endpoint-dipole Green response for one elementary family-shift link. The compact harmonic sum fixes the geometric factor $\gamma_{C_3}=2\lambda_B/9$, while $\lambda_B$ remains an undetermined matching coefficient, and the determinant anisotropy gives $\theta_\ell^{(0)}=-2\lambda_B/9$. Thus $\lambda_B=1$ yields $\theta_\ell^{(0)}=-2/9$ only as a unit-normalized benchmark, not as an exact first-principles prediction. With the current pole masses, this benchmark requires a relative finite matching correction of $9.83\times10^{-6}$ in $m_e/m_\mu$; no microscopic theoretical uncertainty for that correction is derived here.
\end{abstract}

\maketitle
\newpage

\section{Introduction}

The masses of the charged leptons are accurately measured, but their hierarchy is not explained within the Standard Model.  In conventional notation the charged-lepton Yukawa interaction is
\begin{equation}
 {\cal L}_{Y,e}=-\overline L_i (Y_e)_{ij}H E_j+\mathrm{h.c.},
 \label{eq:SMYukawa}
\end{equation}
where $L_i$ is a left-handed lepton doublet of family $i$, $E_j$ is a right-handed charged-lepton singlet of family $j$, $H$ is the Higgs doublet, and $Y_e$ is a dimensionless $3\times3$ matrix in family space.  After electroweak symmetry breaking, in unitary gauge,
\(H\to (0,(v+h)/\sqrt2)^T\), with \(v\simeq 246\,\mathrm{GeV}\)
the Higgs vacuum expectation value and \(h\) the physical Higgs field.
The charged-lepton mass matrix is then \(M_e=vY_e/\sqrt2\) \cite{Glashow1961,EnglertBrout1964,Higgs1964,Guralnik1964,Weinberg1967,Salam1968,PDG2026}.  The Higgs mechanism supplies the electroweak scale, while the eigenvalues of $Y_e$ remain empirical inputs.

Koide observed that the charged-lepton pole masses approximately obey \cite{Koide1981,Koide1983PRD,Koide1983PLB,Koide1990}
\begin{equation}
 Q_\ell\equiv
 \frac{m_e+m_\mu+m_\tau}
 {(\sqrt{m_e}+\sqrt{m_\mu}+\sqrt{m_\tau})^2}
 \simeq \frac23 .
 \label{eq:Koide}
\end{equation}
The striking feature of this relation is the appearance of square roots of masses.  Foot gave the simplest geometric interpretation \cite{Foot1994}: if
\begin{equation}
 \bm z=(\sqrt{m_e},\sqrt{m_\mu},\sqrt{m_\tau}),\qquad
 \hat{\bm d}=\frac{1}{\sqrt3}(1,1,1),
\end{equation}
then Eq.~\eqref{eq:Koide} says that the angle between $\bm z$ and the democratic direction $\hat{\bm d}$ is $\pi/4$.  A large literature has explored Koide-like relations, family-Higgs potentials, family-gauge symmetries, and possible extensions to neutrino and quark sectors \cite{RiveroSponer2005,Sumino2009PLB,Sumino2009JHEP,Koide1983PLB,Koide2010,RodejohannZhang2011,Kartavtsev2012,Cao2012,Zenczykowski2014, GERARD2006563}.  These works suggest that the $\pi/4$ angle to the democratic direction is geometrically robust, whereas fixing the vector's position around that direction on the Koide cone is the harder problem.

The $Z_3$-symmetric square-root-mass parametrization and the numerical proximity of the charged-lepton phase to $2/9$ predate the present work. In particular, {\.Z}enczykowski explicitly discussed $\delta_L\simeq2/9$ and the associated charged-lepton mass pattern, building on earlier observations cited there \cite{Zenczykowski2012,Zenczykowski2014}. Up to cyclic ordering and the sign convention for the phase, this is the same empirical charged-lepton orientation.  
The novelty claimed here is therefore not the number $2/9$ itself, nor the corresponding Koide-based estimate of $m_\tau$. It is the proposed combination of (i) an antiperiodic two-mode seed, (ii) its periodic symmetric-square carrier, (iii) a coherent branch-real order parameter realizing the Koide cone, (iv) a quadratic overlap map to Yukawa eigenvalues, and (v) an endpoint-dipole Green/Berry matching prescription for the residual family shift. The present work combines these ingredients into one compact-cycle effective construction rather than proposing a new empirical mass formula. 


This paper develops a compact-cycle effective realization of the charged-lepton Koide geometry. Its purpose is to show how a vector of square-root masses can arise from a single compact internal amplitude, and how the remaining angle on the Koide cone can be related to a compact Green response.

The starting point is the pair of lowest antiperiodic modes on an internal circle, \(\xi_\pm(\varphi)=e^{\pm i\varphi/2}\). Although each mode changes sign after one turn, their symmetric products are periodic and span\(\{e^{i\varphi},1,e^{-i\varphi}\}\), a minimal three-component family space.
A neutral symmetric coefficient matrix weights these products and defines an internal family amplitude \(Z(\varphi)\). A reality condition and the requirement that this matrix be the square of one underlying two-component spinor give \(s^2=2|d|^2\), which is precisely the Koide \(45^\circ\) cone after sampling at three equally spaced points.


The physical masses are not taken to be linear samples of \(Z\). The ordinary charged-lepton Yukawa matrix is obtained from an overlap of left- and right-handed internal profiles, so that in the tight \(C_3\) limit \(m_a\propto |Z(2\pi a/3)|^2\).


The remaining orientation on the cone is parameterized by treating one $C_3$ family shift as a displacement by $2\pi/3$ on the full internal circle. 
The compact harmonic sum fixes the geometric factor in $\gamma_{C_3}=2\lambda_B/9$, while $\lambda_B$ remains an undetermined matching coefficient. The unit-normalized benchmark $\lambda_B=1$ then gives $\theta_\ell^{(0)}=-2/9$; its normalization is not derived from a microscopic ultraviolet model.

The paper is organized as follows. Sections~\ref{sec:AP}--\ref{sec:selfdual} construct the compact-cycle family amplitude from the lowest antiperiodic doublet and show how its coherent symmetric square gives the Koide cone. Sections~\ref{sec:C3}--\ref{sec:greendress} introduce the residual \(C_3\) clock--shift structure and compute the Green-function dressing of one elementary family shift on the full internal circle. Sections~\ref{sec:Yukawa}--\ref{sec:det} connect this dressed compact-cycle structure to charged-lepton Yukawa eigenvalues and show how the determinant term fixes the remaining orientation on the Koide cone. Sections~\ref{sec:pole} and~\ref{sec:numerics} state the pole-mass interpretation and give the numerical tests. The appendices discuss small departures from the self-dual cone and the stability constraints from higher Fourier harmonics.


\section{Antiperiodic sign doublet and symmetric square}
\label{sec:AP}

We start with an internal circle.  Its coordinate is a dimensionless angle
\begin{equation}
 \varphi\sim\varphi+2\pi .
\end{equation}
The relevant low-lying sector is antiperiodic:
\begin{equation}
\psi(\varphi+2\pi)=-\psi(\varphi).
\label{eq:antiperiodic}
\end{equation}
This is the internal analogue of choosing the antiperiodic spin structure on a
circle. Its Fourier modes carry half-integer internal momentum. The two lowest
modes, with momenta $\pm 1/2$, are
\begin{equation}
 \xi_+(\varphi)=\ee^{+\ii\varphi/2},\qquad
 \xi_-(\varphi)=\ee^{-\ii\varphi/2}.
 \label{eq:lowestdoublet}
\end{equation}
Both change sign under one circuit of the internal circle.  They span a two-dimensional internal sign space,
\begin{equation}
 \Hil_{\rm sign}^{\rm AP}=\mathrm{span}\{\xi_+,\xi_-\}.
 \label{eq:Hsign}
\end{equation}
The word ``sign'' refers to the two nearest branches of the antiperiodic internal spectrum.  It does not denote electric charge, and it is not by itself the particle--antiparticle operation.

The central observation is that the symmetric square of an antiperiodic doublet is periodic.  Explicitly,
\begin{align}
 \xi_+^2&=\ee^{+\ii\varphi},\\
 \sqrt2\,\xi_+\xi_-&=\sqrt2,\\
 \xi_-^2&=\ee^{-\ii\varphi}.
\end{align}
Thus
\begin{equation}
 \mathrm{Sym}^2\Hil_{\rm sign}^{\rm AP}
 =
 \mathrm{span}\{\ee^{+\ii\varphi},1,\ee^{-\ii\varphi}\}.
 \label{eq:sym2basis}
\end{equation}
This gives a minimal three-component periodic carrier. In the effective theory below, this lowest symmetric-square sector is taken as the minimal three-component carrier for the charged-lepton families.

\section{The sign-bilinear order parameter}
\label{sec:Q}

A family profile is encoded by a neutral symmetric sign-bilinear order parameter
\begin{equation}
 Q_{AB}=Q_{BA},\qquad A,B\in\{+,-\}.
 \label{eq:Qdef}
\end{equation}
The field $Q_{AB}$ is neutral with respect to Standard-Model gauge charges.  It is not the Higgs doublet.  It is an internal order parameter that controls the family dependence of the effective charged-lepton Yukawa eigenvalues after the ordinary Higgs field supplies the electroweak scale.

There are two equivalent ways to view $Q$.  First, it may be a collective coordinate of the compact internal defect, specifying how the two antiperiodic sign branches are glued into a periodic family amplitude.  Second, it may be introduced as a Hubbard--Stratonovich field for a neutral sign-bilinear channel in an effective theory of the compact cycle \cite{Hubbard1959,Stratonovich1957}.  In either interpretation, the homogeneous vacuum value of $Q$ determines a family form factor.

For a real charged-lepton mass profile it is convenient to write
\begin{equation}
 Q=
 \begin{pmatrix}
 d & s/\sqrt2\\
 s/\sqrt2 & d^*
 \end{pmatrix},
 \qquad
 s\in\mathbb R,\qquad d=|d|\ee^{\ii\theta}.
 \label{eq:Qmatrix}
\end{equation}
Here $s$ is the singlet component.  It multiplies the middle element of the symmetric square, namely the constant function.  The complex number $d$ is the oriented doublet component.  It multiplies the pair of harmonics $\ee^{+\ii\varphi}$ and $\ee^{-\ii\varphi}$.

The associated family-amplitude form factor on the internal circle is
\begin{align}
 Z(\varphi)
 &=Q_{++}\ee^{\ii\varphi}+\sqrt2 Q_{+-}+Q_{--}\ee^{-\ii\varphi}
 \nonumber\\
 &=s+d\ee^{\ii\varphi}+d^*\ee^{-\ii\varphi}
 \nonumber\\
 &=s+2|d|\cos(\varphi+\theta).
 \label{eq:Zphi}
\end{align}
The form factor $Z(\varphi)$ is dimensionless. It fixes only the family-space shape of the charged-lepton spectrum. Once the common electroweak and overlap normalization is restored, its sampled values determine the normalized square roots of the masses, while the physical masses are proportional to $|Z(\varphi_a)|^2$.

\section{Coherent rank one and the Koide cone}
\label{sec:selfdual}

A generic symmetric $2\times2$ matrix $Q$ contains more freedom than is needed
for the compact-cycle construction. We restrict to the coherent sector, meaning
that the sign-bilinear order parameter is the symmetric square of a single
two-component sign spinor $q_A$:
This coherence condition, together with branch reality below, is an explicit defining assumption of the minimal effective sector; it is not derived from a microscopic vacuum-selection dynamics in the present work.
\begin{equation}
Q_{AB}
=
q_A q_B .
\label{eq:coherent-square}
\end{equation}
For a symmetric $2\times2$ matrix, this condition is equivalent to rank one,
\begin{equation}
\det Q
=
0 .
\label{eq:rank-one-condition}
\end{equation}
This is the algebraic form of coherence. 
For the branch-real parametrization in Eq.~\eqref{eq:Qmatrix}, one has
\begin{equation}
\det Q
=
|d|^2-\frac{s^2}{2}.
\label{eq:det-Q-sd}
\end{equation}
Thus the coherent rank-one condition gives
\begin{equation}
s^2
=
2|d|^2 .
\label{eq:selfdual-condition}
\end{equation}
We will call this the self-dual condition.

The same condition has a more microscopic spinor-level form. Suppose the two
components of the underlying sign spinor are related by branch reality,
\begin{equation}
q_-
=
q_+^*,
\qquad
q_+
=
r e^{i\alpha}.
\label{eq:branch-reality-spinor}
\end{equation}
Then the coherent square gives
\begin{equation}
Q_{++}
=
r^2 e^{2i\alpha}
=
d,
\qquad
Q_{--}
=
r^2 e^{-2i\alpha}
=
d^*,
\qquad
Q_{+-}
=
r^2
=
\frac{s}{\sqrt2}.
\label{eq:branch-real-components}
\end{equation}
Consequently,
\begin{equation}
|d|
=
r^2,
\qquad
s
=
\sqrt2\,r^2,
\label{eq:branch-real-s-d}
\end{equation}
which again gives Eq.~\eqref{eq:selfdual-condition}. Thus branch reality is not
a second independent derivation after rank one. Rather, it is the spinor-level
realization of the same coherent rank-one condition, and it reproduces the real form factor defined in Eq.~\eqref{eq:Zphi}.
Within the assumed coherent branch-real sector, the self-dual condition follows kinematically.

We now connect this self-dual condition to Koide geometry. Write
\begin{equation}
Z(\varphi)
=
s\left[1+\kappa\cos(\varphi+\theta)\right],
\qquad
\kappa
=
\frac{2|d|}{s}.
\label{eq:Z-kappa-form}
\end{equation}
Equation~\eqref{eq:selfdual-condition} is equivalent to
\begin{equation}
\kappa
=
\sqrt2 .
\label{eq:kappa-sqrt-two}
\end{equation}
Equivalently, in the inner product
\begin{equation}
\langle f,g\rangle
=
\int_0^{2\pi}\frac{d\varphi}{2\pi}\,
f^*(\varphi)g(\varphi),
\label{eq:circle-inner-product}
\end{equation}
the singlet part of $Z$ has norm $s^2$, while the oscillating doublet part
$d e^{i\varphi}+d^*e^{-i\varphi}$ has norm $2|d|^2$. Thus
Eq.~\eqref{eq:selfdual-condition} says that the singlet and doublet pieces of
the compact family amplitude have equal norm.

Sampling at three equally spaced phases,
\begin{equation}
\varphi_a
=
\frac{2\pi a}{3},
\qquad
a=0,1,2,
\label{eq:three-sampling-points}
\end{equation}
gives the normalized sampled amplitude
\begin{equation}
z_a
\equiv
\frac{Z(\varphi_a)}{s}
=
1+\sqrt2\cos\left(\theta+\frac{2\pi a}{3}\right).
\label{eq:sampled-root-vector}
\end{equation}
On a branch where the three sampled amplitudes are positive, the vector
$(z_0,z_1,z_2)$ is proportional to the physical square-root mass vector. More
generally, it is the signed square-root vector, while the physical masses remain
proportional to $|Z(\varphi_a)|^2$.

Using
\begin{equation}
\sum_{a=0}^{2}
\cos\left(\theta+\frac{2\pi a}{3}\right)
=
0,
\qquad
\sum_{a=0}^{2}
\cos^2\left(\theta+\frac{2\pi a}{3}\right)
=
\frac{3}{2},
\label{eq:C3-trig-identities}
\end{equation}
one obtains
\begin{equation}
\sum_a z_a
=
3,
\qquad
\sum_a z_a^2
=
6.
\label{eq:sampled-vector-sums}
\end{equation}
Therefore, on the positive square-root branch,
\begin{equation}
\frac{\sum_a m_a}{\left(\sum_a \sqrt{m_a}\right)^2}
=
\frac{\sum_a z_a^2}{\left(\sum_a z_a\right)^2}
=
\frac{6}{9}
=
\frac{2}{3}.
\label{eq:Koide-from-selfdual-samples}
\end{equation}
Equivalently, the square-root mass vector has equal norm along the democratic axis and orthogonal to it. Thus the assumed coherent branch-real sector realizes the $45^\circ$ Koide cone. Since rank one is algebraically equivalent here to the equal-norm condition, this is an effective realization rather than a first-principles derivation of why the physical vacuum satisfies Koide's relation.

\section{Residual \texorpdfstring{$C_3$}{C3} clock and shift}
\label{sec:C3}

The family space used for the charged leptons is the three-dimensional carrier
\begin{equation}
 \Hil_f=\mathrm{span}\{|0\rangle,|1\rangle,|2\rangle\}.
\end{equation}
The $C_3$ clock operator is defined by
\begin{equation}
 \Vhat|a\rangle=\omega^a|a\rangle,\qquad
 \omega=\ee^{2\pi\ii/3}.
 \label{eq:clock}
\end{equation}
The eigenstates of $\Vhat$ are the charged-lepton basis states.  We use the ordering
\begin{equation}
 |0\rangle=|\tau\rangle,\qquad |1\rangle=|\mu\rangle,\qquad |2\rangle=|e\rangle.
 \label{eq:ordering}
\end{equation}
The eigenvalue $\omega^a$ corresponds to the point $\varphi_a=2\pi a/3$ on the internal circle.

The shift operator moves from one clock state to the next.  The bare shift is
\begin{equation}
 \Uhat_0|a\rangle=|a+1\rangle,\qquad |a+3\rangle=|a\rangle,
\end{equation}
and obeys
\begin{equation}
 \Uhat_0\Vhat=\omega^{-1}\Vhat\Uhat_0,\qquad \Uhat_0^3=\mathbf 1.
 \label{eq:bareweyl}
\end{equation}
This is the standard finite clock--shift algebra.  The important point in the present model is that $\Uhat_0$ is not treated as an isolated discrete operation.  It is the projection of a displacement by
\begin{equation}
 \Delta=\frac{2\pi}{3}
\end{equation}
inside the full internal circle.  The full circle contains infinitely many harmonics.  When the theory is reduced to the three-dimensional charged-lepton sector, the discarded harmonics can dress the effective shift operator.  The next section computes this dressing.

\section{Green-dressed shift from a compact Berry connection}
\label{sec:greendress}

A $C_3$ family shift is the projection of a displacement on the full internal
circle. One elementary step moves a clock point by $\Delta=2\pi/3$.
Before the projection to the three-dimensional family carrier, the compact
circle contains all Fourier harmonics. The nonzero harmonics can therefore
dress the elementary shift operator. We describe this dressing by a compact
Berry connection $a(\varphi)$ on the internal circle. This is the standard
slow--fast adiabatic setting: after fast internal modes are eliminated,
transport of the slow family frame is encoded by a Berry connection
\cite{BornOppenheimer1927,Simon1983,Berry1984,WilczekZee1984}.

The fast compact sector is taken to have the canonical phase-field stiffness
\begin{equation}
K
=
-\partial_\varphi^2 .
\label{eq:berry-kernel}
\end{equation}
On zero-average functions its inverse is the circle Green function,
\begin{equation}
G
=
K^{-1},
\qquad
G e^{in\varphi}
=
\frac{1}{n^2}e^{in\varphi},
\qquad
n\neq 0 .
\label{eq:circle-green-function}
\end{equation}

An elementary oriented link compares the two endpoints of the displacement.
We encode it by the neutral endpoint source
\begin{equation}
j_\Delta(\varphi)
=
\delta_{2\pi}(\varphi)
-
\delta_{2\pi}(\varphi-\Delta),
\label{eq:endpoint-dipole-source}
\end{equation}
where $\delta_{2\pi}$ is normalized with respect to $d\varphi/(2\pi)$. Its
Fourier components are
\begin{equation}
(j_\Delta)_n
=
1-e^{-in\Delta}.
\label{eq:endpoint-dipole-fourier}
\end{equation}
The factor $1-e^{-in\Delta}$ measures the response of the $n$th harmonic to
one family step.

At the level of the compact-cycle effective theory, the one-link matching is
specified by the linear response of the fast Berry sector to the endpoint
dipole source. Keeping only the leading Gaussian response of the nonzero
Fourier harmonics gives the most economical translation-invariant functional
\begin{equation}
S_B[a;j_\Delta]
=
-\kappa_B\langle a,Ka\rangle
+
g_B\langle j_\Delta,a\rangle .
\label{eq:general-berry-functional}
\end{equation}

Here $a(\varphi)$ is the compact Berry-response field, $g_B$ is the endpoint charge of one elementary link, and $\kappa_B$ fixes the quadratic normalization of the fast compact sector. We use the inner product in Eq.~\eqref{eq:circle-inner-product} and keep both coefficients arbitrary. Completing the square and dividing by the zero-source Gaussian gives
\begin{equation}
\frac{
\int {\cal D}a\,\exp\{iS_B[a;j_\Delta]\}
}{
\int {\cal D}a\,\exp\{iS_B[a;0]\}
}
=e^{i\gamma(\Delta)},
\qquad
\gamma(\Delta)
=
\frac{g_B^2}{4\kappa_B}
\langle j_\Delta,Gj_\Delta\rangle .
\label{eq:berry-phase-general}
\end{equation}
Using $(j_\Delta)_n=1-e^{-in\Delta}$, this becomes
\begin{equation}
\langle j_\Delta,Gj_\Delta\rangle
=4\sum_{n=1}^{\infty}\frac{1-\cos(n\Delta)}{n^2},
\qquad
\gamma(\Delta)
=
\frac{\lambda_B}{\pi^2}
\sum_{n=1}^{\infty}\frac{1-\cos(n\Delta)}{n^2},
\qquad
\lambda_B\equiv\frac{\pi^2g_B^2}{\kappa_B}.
\label{eq:lambdaB}
\end{equation}
For one $C_3$ step, $\Delta=2\pi/3$, harmonics divisible by three are insensitive to the displacement, and
\begin{equation}
\sum_{n=1}^{\infty}\frac{1-\cos(2\pi n/3)}{n^2}
=\frac{2\pi^2}{9},
\qquad
\gamma_{C_3}=\frac{2\lambda_B}{9},
\qquad
\delta_f=3\gamma_{C_3}=\frac{2\lambda_B}{3}.
\label{eq:general-holonomy}
\end{equation}
The compact Green sum therefore fixes the geometric coefficient $2/9$, but it does not determine the matching coefficient $\lambda_B$. A microscopic completion would have to determine the endpoint charge, the stiffness normalization, possible massive-mode contributions, and any independent bare holonomy. In the numerical discussion below, $\lambda_B=1$ is used only as a unit-normalized benchmark. The matched shift is
\begin{equation}
\hat U_{\rm eff}=e^{i\gamma_{C_3}}\hat U_0,
\qquad
\hat U_{\rm eff}^{\,3}=e^{i\delta_f}\mathbf 1 .
\label{eq:matched-shift-general}
\end{equation}

\section{Quadratic Yukawa overlap}
\label{sec:Yukawa}

We now connect the compact-cycle family amplitude to the charged-lepton
Yukawa eigenvalues. The amplitude $Z(\varphi)$ becomes an operator on the
three-dimensional clock space by replacing the phase $e^{i\varphi}$ with the
clock operator $\hat V$:
\begin{equation}
Z(\hat V)
=
s\mathbf 1
+
d\hat V
+
d^*\hat V^\dagger .
\label{eq:Z-clock-operator}
\end{equation}
Using the clock eigenstates in Eq.~\eqref{eq:clock}, the operator $Z(\hat V)$ samples the compact amplitude at the three clock points:
\begin{equation}
Z(\hat V)|a\rangle
=
Z(\varphi_a)|a\rangle,
\qquad
\varphi_a
=
\frac{2\pi a}{3}.
\label{eq:Z-clock-samples}
\end{equation}

The physical masses are not assumed to be linear samples of $Z$. In the
four-dimensional effective theory, the charged-lepton Yukawa matrix arises from
an overlap of left- and right-handed internal profiles. To make this explicit,
write the internal-circle extensions of the Standard-Model lepton doublet and
charged-lepton singlet as
\begin{equation}
L(x,\varphi)
=
\sum_a L_a(x) f^L_a(\varphi),
\qquad
E(x,\varphi)
=
\sum_b E_b(x) f^R_b(\varphi).
\label{eq:internal-profile-expansion}
\end{equation}
A local Yukawa interaction on the compact cycle has the form
\begin{equation}
\mathcal L_Y
=
-y_0
\int_0^{2\pi}\frac{d\varphi}{2\pi}\,
L(x,\varphi)H(x)E(x,\varphi)
+
{\rm h.c.}
\label{eq:local-internal-yukawa}
\end{equation}
and gives the family-space Yukawa matrix
\begin{equation}
(Y_\ell)_{ab}
=
y_0
\int_0^{2\pi}\frac{d\varphi}{2\pi}\,
[f^L_a(\varphi)]^*
f^R_b(\varphi).
\label{eq:yukawa-overlap}
\end{equation}

In the aligned compact-family sector the left- and right-handed profiles carry
the same family amplitude:
\begin{equation}
f^L_a(\varphi)
=
Z(\varphi)w_a(\varphi),
\qquad
f^R_a(\varphi)
=
Z(\varphi)w_a(\varphi).
\label{eq:aligned-profiles}
\end{equation}
Here $w_a(\varphi)$ is a clock-localized profile centered at
$\varphi_a=2\pi a/3$. In the tight $C_3$ limit,
\begin{equation}
w_a^*(\varphi)w_b(\varphi)
\simeq
\delta_{ab}\,\delta_{2\pi}(\varphi-\varphi_a),
\label{eq:tight-C3-limit}
\end{equation}
where $\delta_{2\pi}$ is normalized with respect to $d\varphi/(2\pi)$. The
Yukawa matrix is then diagonal in the clock basis:
\begin{equation}
(Y_\ell)_{ab}
\simeq
y_0 |Z(\varphi_a)|^2 \delta_{ab}.
\label{eq:yukawa-diagonal-samples}
\end{equation}
Equivalently,
\begin{equation}
Y_\ell
=
y_* Z^\dagger(\hat V)Z(\hat V),
\label{eq:quadratic-yukawa-operator}
\end{equation}
up to the common normalization $y_*$, which includes the overlap normalization
of the localized profiles.

In the minimal construction $H(x)$ is the ordinary four-dimensional Standard-Model Higgs doublet and carries no dependence on the internal coordinate $\varphi$. Electroweak symmetry breaking therefore multiplies all three Yukawa eigenvalues by the common factor $v/\sqrt{2}$ but does not determine their ratios or the orientation on the Koide cone. The minimal truncation contains no direct Higgs--order-parameter portal, for example $c_{HQ}(H^\dagger H)\Tr(Q^\dagger Q)$, and backreaction of electroweak symmetry breaking on the compact geometry is neglected. Such portal terms or Higgs-induced deformations of the localized profiles would contribute to the finite matching corrections discussed in Sec.~\ref{sec:numerics}.

After electroweak symmetry breaking, the charged-lepton masses are therefore
\begin{equation}
m_a
=
\frac{v}{\sqrt 2}\,
y_* |Z(\varphi_a)|^2 .
\label{eq:masses-from-Z-samples}
\end{equation}
Using Eqs.~\eqref{eq:Z-kappa-form} and \eqref{eq:kappa-sqrt-two}, one obtains the one-scale spectrum
\begin{equation}
m_a
=
M
\left[
1+\sqrt 2\cos\left(\frac{2\pi a}{3}+\theta\right)
\right]^2,
\qquad
a=0,1,2,
\label{eq:general-selfdual-spectrum}
\end{equation}
with
\begin{equation}
M
=
\frac{v}{\sqrt 2}\,y_*s^2 .
\label{eq:mass-scale-M}
\end{equation}
At this stage the Koide cone has fixed the relative singlet and doublet weights
in $Z$, but it has not fixed the remaining orientation angle $\theta$ on the
cone. The next section shows how the Green-dressed compact shift selects this
angle.

\section{Determinant anisotropy}
\label{sec:det}
We now use the matched link and the three-link holonomy in Eqs.~\eqref{eq:general-holonomy} and \eqref{eq:matched-shift-general}.
The determinant term is built from the compact-cycle family amplitude in the
shift sector. To avoid confusing this object with the physical Yukawa matrix
$Y_\ell$, we denote it by
\begin{equation}
\mathcal Z_{\delta_f}
=
s\mathbf 1
+
d\hat U_{\rm eff}
+
d^*\hat U_{\rm eff}^\dagger .
\label{eq:Zdeltaf-definition}
\end{equation}
This is the shift-space analogue of the clock-space amplitude $Z(\hat V)$. The
clock-space amplitude gives the Yukawa eigenvalues through the quadratic overlap
of Sec.~\ref{sec:Yukawa}; the shift-space amplitude controls the phase-sensitive
determinant anisotropy.

Using $\hat U_{\rm eff}^3=e^{i\delta_f}\mathbf 1$, one finds
\begin{equation}
\det \mathcal Z_{\delta_f}
=
s^3
-
3s|d|^2
+
e^{i\delta_f}d^3
+
e^{-i\delta_f}d^{*3}.
\label{eq:det-Zdeltaf}
\end{equation}
The first two terms are independent of the orientation of $d$. The last two
terms depend on the phase
\begin{equation}
d
=
|d|e^{i\theta}
\label{eq:d-phase-theta}
\end{equation}
and therefore select the remaining angle on the Koide cone.

A symmetry-compatible effective potential may be written as
\begin{equation}
V_{\rm eff}(Q)
=
V_{\rm radial}(Q)
+
\lambda_{\rm SD}|\det Q|^2
-
\rho\,{\rm Re}\,\det \mathcal Z_{\delta_f}
+
\cdots .
\label{eq:effective-potential-det}
\end{equation}
Here $V_{\rm radial}$ stabilizes the overall magnitude of $Q$, while
$\lambda_{\rm SD}>0$ penalizes departures from the coherent rank-one cone. The $\lambda_{\rm SD}$ term is a phenomenological penalty implementing the assumed coherent sector; it is not used here as a microscopic derivation of rank one.
The orientation-dependent part is
\begin{equation}
V_\theta
=
-2\rho |d|^3 \cos(3\theta+\delta_f).
\label{eq:theta-potential}
\end{equation}
For $\rho>0$ the minimum satisfies
\begin{equation}
3\theta+\delta_f=0
\qquad
{\rm mod}\;2\pi .
\label{eq:theta-minimum-condition}
\end{equation}
Thus, in the physical chamber used for the charged-lepton ordering, the leading angle is
\begin{equation}
\theta_\ell^{(0)}=-\frac{\delta_f}{3}=-\gamma_{C_3}=-\frac{2\lambda_B}{9}.
\label{eq:theta-lepton-general}
\end{equation}
Thus the determinant anisotropy relates the orientation on the Koide cone to the Berry-sector matching coefficient. The effective construction gives the relation $\theta_\ell^{(0)}=-2\lambda_B/9$; the numerical value $-2/9$ follows only for the unit-normalized benchmark $\lambda_B=1$.

For the benchmark $\lambda_B=1$, Eq.~\eqref{eq:general-selfdual-spectrum} is evaluated at $\theta=-2/9$.
The same result can be viewed as a dressed three-link vertex. The term $d^3$ corresponds to the oriented process $0\to 1\to 2\to 0$
around the family triangle. Each elementary link carries the Berry phase $\gamma_{C_3}$, so the determinant vertex carries the total phase $\delta_f=3\gamma_{C_3}$. 
This three-link phase gives $\theta_\ell^{(0)}=-2\lambda_B/9$; the unit-normalized benchmark $\lambda_B=1$ gives $-2/9$.

\section{Pole-mass interpretation}
\label{sec:pole}

Koide's relation is conventionally most accurate for pole masses, and the present construction is therefore formulated as an on-shell effective matching condition,
\begin{equation}
Y_\ell^{\rm pole}\equiv\frac{\sqrt2}{v}M_\ell^{\rm pole}=y_*Z^\dagger(\hat V)Z(\hat V).
\label{eq:poleY}
\end{equation}
This is an assumption about infrared matching, not a demonstrated renormalization-group invariant relation. Explicit running-mass analyses find that the charged-lepton Koide parameter at $\mu=M_Z$ differs from its pole-mass value by approximately $0.2\%$ \cite{XingZhang2006}, about two hundred times the $9.83\times10^{-6}$ benchmark mass-ratio residual discussed below, although the two numbers refer to different observables. We therefore treat the compact relation strictly as an infrared boundary condition. The present work does not provide the symmetry or threshold mechanism needed to protect or reproduce it at higher scales; such a mechanism is required in a UV completion, as illustrated by Sumino-type approaches \cite{Sumino2009PLB,Sumino2009JHEP}.

\section{Numerical tests}
\label{sec:numerics}

We use the 2026 Particle Data Group pole masses \cite{PDG2026},
\begin{equation}
\begin{aligned}
m_e&=0.51099895069\pm0.00000000016~{\rm MeV},\\
m_\mu&=105.6583755\pm0.0000023~{\rm MeV},\\
m_\tau&=1776.93\pm0.09~{\rm MeV}.
\end{aligned}
\label{eq:fit-input-masses}
\end{equation}
The ratio $m_e/m_\mu$ fixes $\theta$ independently of the common scale in Eq.~\eqref{eq:general-selfdual-spectrum}. Treating the quoted mass uncertainties as independent and using standard linear propagation, $\sigma_{\ln r}^2=(\sigma_e/m_e)^2+(\sigma_\mu/m_\mu)^2$, gives
\begin{equation}
\theta_{e\mu}=-0.222222047074\pm3.88\times10^{-10},
\qquad
\theta_{e\mu}+\frac29=(1.75148\pm0.00388)\times10^{-7}.
\label{eq:theta-error-compact}
\end{equation}
With experimental uncertainties alone, the offset in Eq.~\eqref{eq:theta-error-compact} has the formal significance $451.7\sigma$. Since $\lambda_B=1$ is only a benchmark and the present effective theory does not assign a microscopic matching uncertainty, this number should not be interpreted as a statistical exclusion of the compact-cycle framework. Instead, it fixes the finite correction that a microscopic completion must generate:
\begin{equation}
\frac{(m_e/m_\mu)_{\lambda_B=1}-(m_e/m_\mu)_{\rm exp}}
{(m_e/m_\mu)_{\rm exp}}
=-9.8334\times10^{-6}.
\label{eq:ratio-residual}
\end{equation}
The requested weighted fit minimizes $\chi^2=\sum_a[m_a^{\rm exp}-M f_a(\theta)]^2/\sigma_a^2$, with $f_a$ defined by Eq.~\eqref{eq:general-selfdual-spectrum}. The fit and the two tau-mass comparisons are summarized in Table~\ref{tab:tests}.

\begin{table}[t]
\caption{ Compact numerical summary. Parentheses denote propagated experimental uncertainties only; no theoretical matching uncertainty is assigned.}
\label{tab:tests}
\begin{ruledtabular}
\begin{tabular}{lcc}
Test &  Quantity &  Result \\
\hline
$m_e/m_\mu$ & $\theta_{e\mu}$ & $-0.222222047074(388)$\\
weighted $(M,\theta)$ fit & $M_{\rm fit}$ & $313.8564003(63)~{\rm MeV}$\\
weighted $(M,\theta)$ fit & $\theta_{\rm fit}$ & $-0.222222047074(388)$\\
weighted $(M,\theta)$ fit & $\chi^2/{\rm d.o.f.}$ & $0.188/1$\\
equivalent Berry coefficient & $\lambda_B^{\rm fit}$ & $0.9999992118(17)$\\
cone from $m_e,m_\mu$ & $m_\tau$ (pull) & $1776.969027(35)~{\rm MeV}$ $(0.434\sigma)$\\
$\lambda_B=1$, normalized to $m_\mu$ & $m_\tau$ (pull) & $1776.967497(39)~{\rm MeV}$ $(0.417\sigma)$\\
unit benchmark & $\delta(m_e/m_\mu)/(m_e/m_\mu)$ & $-9.8334\times10^{-6}$\\
\end{tabular}
\end{ruledtabular}
\end{table}

The weighted fit is dominated by $m_e$ and $m_\mu$, while the present uncertainty of $m_\tau$ makes both tau comparisons much weaker tests. Equivalently, the fitted Berry coefficient differs from the unit benchmark by only $\lambda_B^{\rm fit}-1=-7.88\times10^{-7}$. The required correction can also be compared directly with Appendix~\ref{app:higher}: using the linear response in Appendix~\ref{app:perturb}, the offset in Eq.~\eqref{eq:theta-error-compact} corresponds either to an angularly induced electron-amplitude shift $\delta Z_e=1.82\times10^{-7}$ or, if represented entirely as a local electron-node correction, to $\delta Z_e=1.98\times10^{-7}$. Thus
\begin{equation}
|\delta Z_e|\simeq(1.8\text{--}2.0)\times10^{-7},
\label{eq:required-delta-Ze}
\end{equation}
which is of the same order as the sensitivity bound $|\delta Z_e|\lesssim2\times10^{-7}$ in Appendix~\ref{app:higher}. This is a consistency-of-scales statement, not a calculation of the correction or an assigned theoretical error bar.

\section{Discussion}
\label{sec:discussion}

The construction separates the charged-lepton problem into two parts. 
Under the explicit coherent branch-real assumption, the lowest antiperiodic doublet realizes the Koide cone through equal singlet and doublet norms. The compact Berry sector then gives the relation $\theta_\ell^{(0)}=-2\lambda_B/9$ for the remaining orientation; $\lambda_B=1$ is the unit-normalized benchmark, not a microscopic prediction.


The numerical success of the construction relies on two matching assumptions.
First, the compact Berry sector is evaluated at the unit-normalized benchmark
$\lambda_B=1$ of Eq.~\eqref{eq:lambdaB}. A microscopic completion
must explain why additional compact charges, massive fast modes, or a bare
link holonomy do not appreciably shift this value. Second, the matching is
interpreted at the pole-mass level, as in Eq.~\eqref{eq:poleY}. This
requires either a UV mechanism that reproduces the compact-cycle relation for
pole masses or a protection mechanism for the corresponding running relation.

The near-node origin of the electron mass makes higher-harmonic stability a
sharp constraint. Since $Z_e$ is small at the selected angle, even a tiny
higher-harmonic correction to the compact amplitude can produce an enhanced
relative shift in $m_e$. Appendix~\ref{app:higher} shows that
harmonics beyond $k=0,\pm1$ must therefore be symmetry-forbidden or strongly
suppressed if the observed $10^{-5}$-level accuracy of the charged-lepton
ratios is to remain stable.

The present work treats $Q$ only through its homogeneous background value. To make the EFT limitation explicit, suppose a microscopic extension contains propagating Standard-Model-singlet fluctuations $q_r$. Their leading gauge-invariant coupling can be parameterized as
\begin{equation}
{\cal L}_{q\ell}=-\overline L_i H E_j\left[(Y_\ell^{(0)})_{ij}+\sum_r\frac{(C_r)_{ij}}{\Lambda_Q}q_r\right]+{\rm h.c.}
\label{eq:q-fluctuation-coupling}
\end{equation}
After electroweak symmetry breaking, $g_{ij}^{(r)}=v(C_r)_{ij}/(\sqrt2\Lambda_Q)$. Integrating out a mode of mass $m_r$ gives the schematic low-energy coefficients
\begin{equation}
C_{4\ell}^{ijkl}\sim\sum_r\frac{g_{ij}^{(r)}g_{kl}^{(r)*}}{m_r^2},
\qquad
C_{e\gamma}^{ij}\sim\frac{e}{16\pi^2}\sum_{r,k}\frac{g_{ik}^{(r)}g_{jk}^{(r)*}m_k}{m_r^2}.
\label{eq:lfv-power-counting}
\end{equation}
Exact alignment means that every $C_r$ is diagonal in the same basis as $Y_\ell^{(0)}$, so the leading off-diagonal coefficients vanish. Misalignment is constrained by searches for $\mu^+\to e^+\gamma$ and lepton-flavor-violating tau decays \cite{MEG2016,Belle2010,BaBar2010}. The present construction specifies neither $m_r$, $\Lambda_Q$, nor the matrices $C_r$, and therefore cannot predict branching fractions. In the minimal aligned background the Standard-Model charged-current vertex is unchanged at tree level and the SMEFT operator $O_{H\ell}^{(3)}$ is not generated at tree level. A dynamical extension may generate it radiatively, so charged-current lepton-universality measurements provide an additional model-dependent constraint.


The main theoretical task is therefore to embed the compact-cycle effective description in a microscopic antiperiodic Dirac/Yukawa model. Such a completion should reproduce the three essential low-energy ingredients identified here: a branch-real coherent symmetric square giving the Koide cone, a dynamically specified endpoint coupling and stiffness normalization for one elementary family shift, and an inverse-Laplacian compact response with controlled additional holonomy contributions.

\textit{Acknowledgments.—}
This work was partially supported by JSPS KAKENHI Grant Number 23H04864.

\appendix

\section{Small deformations away from the self-dual cone}
\label{app:perturb}

To parametrize small departures from the ideal self-dual spectrum, write
\begin{equation}
 s^2=2|d|^2(1+\eta),\qquad
 \theta=\theta_\ell^{(0)}+\Delta\theta,\quad \theta_\ell^{(0)}=-\frac{2\lambda_B}{9}.
\end{equation}
Then
\begin{equation}
 \beta=\frac{2|d|}{s}=
 \sqrt{\frac{2}{1+\eta}}
 =
 \sqrt2\left(1-\frac{\eta}{2}+O(\eta^2)\right),
\end{equation}
and
\begin{equation}
 m_a=Mx_a^2,\qquad
 x_a=1+\beta\cos\left(\frac{2\pi a}{3}+\theta\right).
\end{equation}
Using the trigonometric identities for three equally spaced points,
\begin{equation}
 \sum_{a=0}^{2}\cos\left(\frac{2\pi a}{3}+\theta\right)=0,\qquad
 \sum_{a=0}^{2}\cos^2\left(\frac{2\pi a}{3}+\theta\right)=\frac32,
\end{equation}
one obtains
\begin{equation}
 Q_\ell=\frac13+\frac{\beta^2}{6}
 =
 \frac23-\frac{\eta}{3}+O(\eta^2).
\end{equation}
Thus the deviation of $Q_\ell$ from $2/3$ measures departure from the self-dual cone.

For the unit-normalized benchmark $\lambda_B=1$, at $\eta=0$ and $\theta=-2/9$, the two independent logarithmic ratios respond as
\begin{align}
 \delta\ln\frac{m_e}{m_\mu}
 &\simeq 23.06\,\eta+56.14\,\Delta\theta,\\
 \delta\ln\frac{m_\mu}{m_\tau}
 &\simeq 1.303\,\eta-4.917\,\Delta\theta,
\end{align}
up to terms quadratic in $\eta$ and $\Delta\theta$.

\section{Higher harmonic stability}
\label{app:higher}

The truncation
\begin{equation}
 Z(\varphi)=s+d\ee^{\ii\varphi}+d^*\ee^{-\ii\varphi}
\end{equation}
is important because the electron sits close to a node.  Suppose higher harmonics shift the amplitude by
\begin{equation}
 \delta Z(\varphi)=c_2\ee^{2\ii\varphi}+c_2^*\ee^{-2\ii\varphi}+\cdots.
\end{equation}
The relative electron-mass shift is approximately
\begin{equation}
 \frac{\delta m_e}{m_e}\simeq 2\frac{\delta Z_e}{Z_e}.
\end{equation}
For the unit-normalized benchmark $\lambda_B=1$, at $\theta=-2/9$, $Z_e\simeq0.04035$. Maintaining relative accuracy at the $10^{-5}$ level therefore requires $|\delta Z_e|\lesssim 2\times10^{-7}$. Any ultraviolet completion must therefore suppress higher harmonics by symmetry or by a strong dynamical hierarchy. The quantitative comparison with the correction required by the pole-mass ratio is given in Sec.~\ref{sec:numerics}; the present estimate remains a sensitivity bound, not a calculation of that correction.

\bibliography{lepton_prd_revision_clean}

\begin{center}
\small
\textcopyright{} 2026 Journal of Physics G: Nuclear and Particle Physics. 
This is the accepted manuscript of the following article:

\medskip

K.~Shulga, ``Charged-lepton Koide geometry from a green-dressed compact family cycle,'' 
\textit{Phys. G: Nucl. Part. Phys.} \textbf{53}, J.  105001 (2026).

\medskip

The final published version is available at 
\href{https://doi.org/10.1088/1361-6471/aea4ce}
{https://doi.org/10.1088/1361-6471/aea4ce}. 
This manuscript version is posted with permission for 
non-commercial scholarly use.
\end{center}

\end{document}